\documentclass[a4paper,11pt]{article}

\usepackage{pos}
\usepackage{graphicx} 
\usepackage{url, hyperref}

\newcommand{\gcn}[1]{\href{https://gcn.nasa.gov/circulars/#1}{GCN {#1}}}

\title{Results from IceCube Searches for High-energy Neutrinos Coincident with Gravitational-Wave Alerts in LVK O4}

\ShortTitle{IceCube GW Follow-up in LVK O4}

\author{The IceCube Collaboration \\{\normalsize \normalfont(a complete list of authors can be found at the end of the proceedings)}\\}

\emailAdd{zsuzsa@astro.columbia.edu}
\emailAdd{jthwaites@icecube.wisc.edu}
\emailAdd{justin.vandenbroucke@wisc.edu}
\emailAdd{veske@metu.edu.tr}

\abstract{
Mergers of compact objects, binary black holes and mergers including at least one neutron star, are a predicted source of high-energy neutrinos. These astrophysical events are now routinely detected through observation of their gravitational wave signature and, at least in one instance, their electromagnetic counterparts were also detected. Particles accelerated during the coalescence of compact objects may also interact to produce high-energy neutrinos, which have yet to be detected, but observations are ongoing. The LIGO-Virgo-KAGRA Collaboration publicly releases information on candidate gravitational wave events from compact binary coalescences in low latency during the current observing run (O4). To aid the electromagnetic follow-up, using data from the IceCube Neutrino Observatory, we search, in real time, for neutrinos spatially and temporally coincident with these gravitational wave candidate events using a time window of 1000 seconds centered on the merger time. We use two methods, both of which have been previously used to search for neutrino emission from gravitational-wave transients: an unbinned maximum likelihood analysis applied to significant alerts and a Bayesian analysis with astrophysical priors, applied to both significant and low-significance alerts. In addition, we search for long-duration neutrino emission up to 14 days after the merging of binaries containing a neutron star. We report analysis results determined in real time for these searches, and set upper limits on both flux and isotropic-equivalent energy emitted in neutrinos.

\vspace{4mm}

{\bfseries Corresponding authors:}

Zsuzsa Marka$^{1*}$,
Jessie Thwaites$^{2}$,
Justin Vandenbroucke$^{2}$,
Do\u{g}a Veske$^{1,3}$\\

{$^{1}$ \itshape Columbia Astrophysics and Nevis Laboratories, Columbia University}\\
{$^{2}$ \itshape Department of Physics and Wisconsin IceCube Particle Astrophysics Center, University of Wisconsin—Madison}\\
{$^{3}$ \itshape Fizik B\"ol\"um\"u, Orta Do\u{g}u Teknik \"Universitesi, \c{C}ankaya, Ankara 06800, Turkey}\\[4mm]
$^*$ Presenter
}

\ConferenceLogo{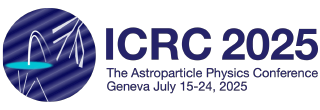}

\FullConference{39th International Cosmic Ray Conference (ICRC2025)\\
 15–24 July 2025\\
Geneva, Switzerland\\}

\begin{document}

\maketitle

\section{Introduction}

Gravitational waves are now routinely detected by LIGO-Virgo-KAGRA (LVK)~\cite{gwtc3}. As messengers, they provide a decipherable fingerprint about the emitting cosmic source. The waveform for every detection by LVK to date can be understood by invoking a compact binary merger model containing neutron stars and black holes. Similarly to gravitational waves, neutrinos travel unobstructed through the Universe, although they do not provide as rich information about the emitting source. Thus, spatial and temporal coincidence-based multimessenger techniques play a key role in source inference.

High energy neutrinos are expected to be produced in mergers of compact objects, either from hadronic interactions in jets formed during the merger \cite{kimura_gwnu}, or from long-lived remnants after the merger \cite{fang_gwnu}. Neutrinos coincident with a detected merger event can additionally help to inform follow-up by smaller field of view instruments. Gravitational wave (GW) events are difficult to localize with GW information alone, especially with a limited number of GW detectors. Typical skymaps sent by LVK often span $\mathcal{O}(100-1,000)$ square degrees, while track-like neutrino events can be localized to $\mathcal{O}(1)$ square degrees \cite{icecube_o3}. A neutrino event coincident with a GW event and shared in real-time can therefore reduce the required search area for electromagnetic follow-up by orders of magnitude. Coordinated multimessenger observations of the sources of GW radiation are needed to understand the processes occurring during the mergers of these objects.

We leverage the all-sky field of view and high uptime of the IceCube Neutrino Observatory, a cubic kilometer detector in the ice at the geographic south pole, to follow up these events. IceCube has followed up GW events from the previous observing runs of the LVK detectors \cite{icecube_o1o2, icecube_o3}. The fourth observing run (O4) of the LVK network started in May 2023 and is ongoing, having been extended multiple times. In this paper, we describe the IceCube Collaboration's real-time gravitational-wave multimessenger follow-up program, its performance measures, demonstrate its action through an example event, and provide cumulative search results up to April 1, 2025 when the LVK detectors stopped for a 2 months long commissioning break.

\section{Search Methods}

The search for neutrino emission coinciding with observation in another messenger is inherently model-dependent. One assumes that the astrophysical source can emit the different messenger types, with specific messenger energy and time characteristics. In the case of a transient astrophysical event characterized by a particular event time, $t_0$, the assumed time dependence of the emission is captured by the search time window. IceCube uses two independent searches (pipelines) in real-time: an astrophysical priors-based method, which runs on both significant and low-significance GW candidates, and a generic one, which runs on the significant GW detections. Both searches run automatically when a new gravitational-wave alert is received via GCN. The searches use a dataset of high energy track events, called the \textit{Gamma-ray Follow-up} (GFU) event stream, available with low latency from the South Pole. This event sample has an average rate of $6-7$ mHz \cite{Abbasi:2021}. We search for coincidences in a $\pm 500$ second time window with both pipelines, motivated by studies of GRB emission \cite{deltat_1000_gw}. 

The \textit{Low-Latency Algorithm for Multimessenger Astrophysics} (LLAMA) pipeline~\cite{bartos_llama_methods,2019arXiv190105486C,stefthesis,icecube_o1o2, icecube_o3} employs model dependence in a Bayesian fashion, assuming astrophysical priors. The astrophysical source is a compact binary coalescence (CBC) event also capable of emitting high-energy neutrinos within a time window of $\pm 500$ seconds from the merger time.
We assume (1) a uniform prior distribution of events within the observation time period; (2) a uniform prior distribution for source sky location; 
(3) an $r^2$ prior distribution up to the GW detection range for source distance; and (4) a log uniform distribution in isotropic emission energies for both messengers. 
The model uses the $\pm500$s search window, which was established by assuming a GRB source model~\cite{deltat_1000_gw}, with a triangular profile to account for increased emission probability close to the merger. 

To find the significance, we first calculate a Bayesian odds ratio using the source model and observational inputs. For GW observations, we use the gravitational-wave skymap with distance information, the candidate event time, a proxy for the signal-to-noise ratio (SNR), and information on the purity of the candidate ($p_{\rm astro}$). From the neutrino detector's side, the inputs are the reconstructed time of arrival, energy, direction, and angular uncertainty.  We find the frequentist significance by comparing the odds ratio for each event to a distribution built empirically from background simulations, using scrambled neutrinos and the gravitational wave candidate skymap set.

The \textit{Unbinned Maximum Likelihood} analysis (UML, also called the ``generic transient'' search in public notices) uses an extended likelihood, used previously in searches for neutrinos from short transients \cite{Braun:2008bg, Braun:2009wp}. This method has been used in previous searches for neutrinos from GW events during O1, O2 \cite{icecube_o1o2}, and O3 \cite{icecube_o3}. It uses the HEALPix pixelization scheme \cite{healpix_methods}, and applies the GW skymap as a penalty to the test statistic (TS) at each location on the sky. The most likely direction given the neutrino and GW data is then computed as the maximum TS pixel, with an associated $p$-value calculated by comparing to background-only pseudo-experiments. 

We also perform an extended time window search using the UML framework, in the case that one or both of the progenitors was likely a neutron star. This search is motivated by some models for extended timescale neutrino emission \cite{fang_gwnu}, and uses a time window of $[-0.1,\, +14]$ days with respect to the merger time. For an event to be run with this time window, it must pass at least one of three criteria: (1) the probability that it is a binary neutron star merger or a neutron star\textendash black hole merger is greater than 50\% ($p_{\mathrm{BNS}}+p_{\mathrm{NSBH}}>0.5$); (2) the probability that the event, if astrophysical, contains a neutron star is greater than 50\% (\texttt{HasNS}$>0.5$); or (3) the probability that some mass was ejected from the system during the merger is greater than 50\% (\texttt{HasRemnant}$>0.5$). To date, 5 events in O4 have passed one or more of these criteria and have been analyzed with this time window.

\section{Real-time Performance}

\begin{figure}
    \centering
    \includegraphics[width=0.55\linewidth]{ 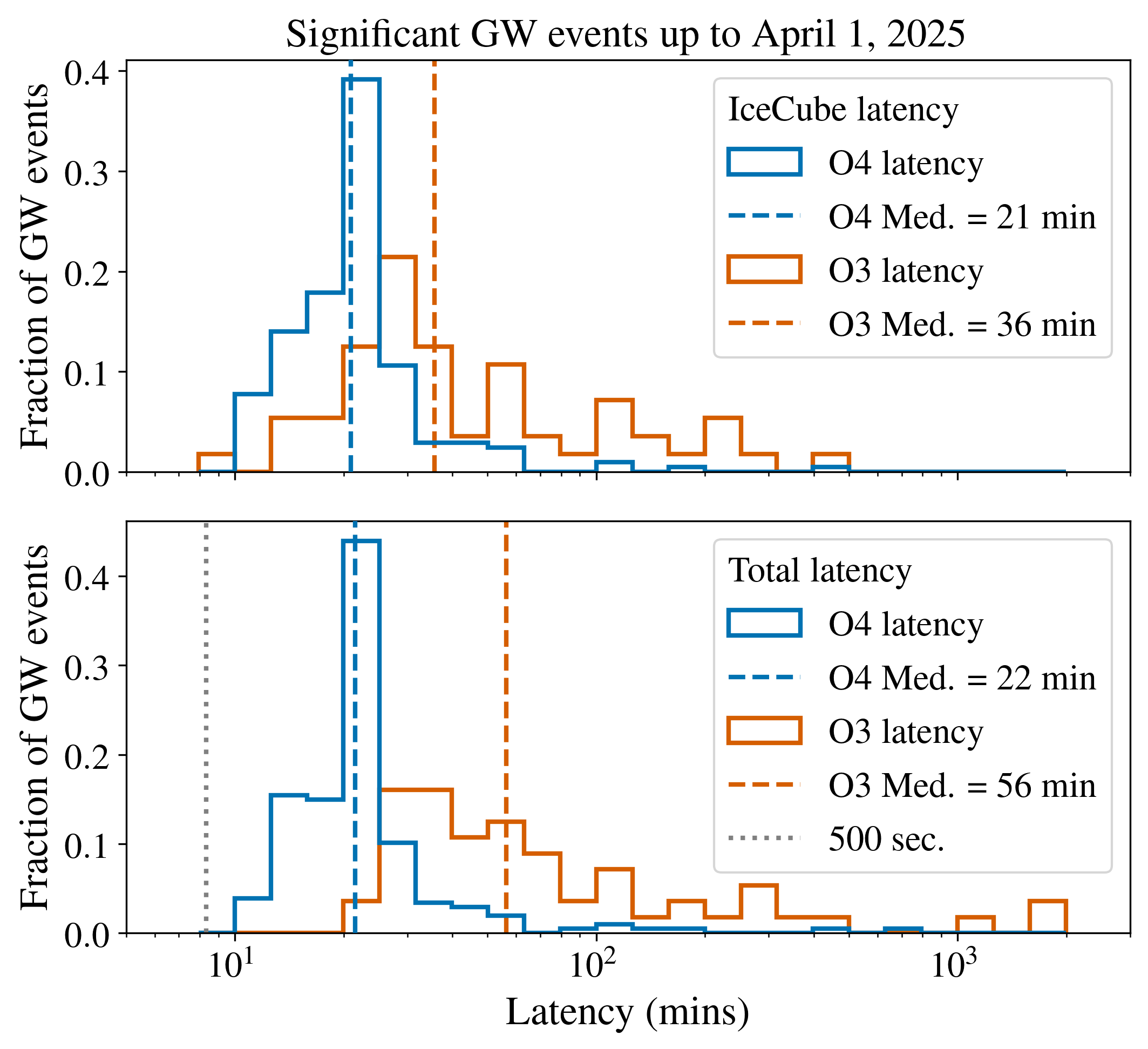}
    \caption{Latency distribution for the joint UML and LLAMA GW follow-up for the high significance LVK alerts. In O3, all events were sent as GCN Circulars, while in O4 most events have been sent automatically via GCN Notice. The top panel shows the IceCube-only latency, while the bottom panel shows the total latency (with reference to the merger time). Only the first GCN for each Significant event in O4 is shown here, and this includes some events at the beginning of O4 sent as circulars before the notice stream was approved.}
    \label{fig:latency}
\end{figure}

IceCube uses the General Coordinates Network (GCN)\footnote{\url{https://gcn.nasa.gov/}} to publish results in real time. During O3, IceCube sent all results manually as GCN Circulars, with human-in-the-loop manual vetting. In O4, we have switched to a fully automated machine-readable GCN Notice stream over Kafka, with the topic \texttt{gcn.notices.icecube.lvk\_nu\_track\_search} \footnote{See documentation hosted at \url{https://gcn.nasa.gov/missions/icecube}}$^{,}$\footnote{Results publicly archived at \url{https://roc.icecube.wisc.edu/public/LvkNuTrackSearch/}}. The switch to this automated system has improved the median response latency from 56 minutes to 22 minutes, from the time that the objects merge, as shown in Figure \ref{fig:latency}. 

Both pipelines run automatically on a separate server upon receiving an LVK alert through GCN, before the available results are combined in an automatic GCN Notice. The largest fraction of the 22-minute latency is waiting for the end of the search time window (500 seconds = 8.3 min). In order not to miss a neutrino, we also wait until we receive the first GFU neutrino after the search time window passes, which can take several minutes due to random arrivals and uncertainty in the latency of the data transfer from the South Pole. LLAMA is very efficient and produces a significance calculation result in a minute~\cite{stefthesis}, except for edge cases. For UML, the latency is on the order of minutes once the time window has closed. LVK publishes multiple alerts for each event as more sophisticated algorithms are used to improve the localization and parameter estimation; the IceCube analyses are performed automatically and a GCN Notice is distributed in response to each update sent by LVK through GCN.

Each Notice reports the $p$-values for the search(es), the time-integrated flux sensitivity range, and the sensitive energy range, assuming an $E^{-2}$ neutrino spectrum, corresponding to the $90\%$ probability region of the GW skymap. For events that reach a $p$-value below $10\%$, we distribute the direction and angular uncertainty for neutrino(s) that reached the threshold in the GCN Notice, to enable prompt electromagnetic follow-up. Joint candidates below a $1\%$ threshold, we also distribute via human-in-the-loop GCN Circulars to further encourage response from the astronomy community. The reported $p$-values are not corrected by any trial factor. 

\section{Selected Results for Individual Events}

In O4, there have been two GW events which had $p<0.01$ in one pipeline which have prompted responses from the community. In both cases, electromagnetic telescopes followed up the identified coincident event(s) sent by IceCube via GCN Circular.

LVK alert S230904n during O4a was a significant event, likely a BBH merger from 1.1~Gpc distance, detected by the Livingston and Hanford LIGO detectors. LLAMA measured a 0.0037 $p$-value, which was distributed to the astronomy community (\gcn{34616}). In response, the identified coincident neutrino was followed up by the Zwicky Transient Facility (ZTF), which found a potential counterpart (\gcn{34717}). The transient was later identified as a supernova type Ia not associated with the GW event (\gcn{34751}). 

\begin{figure}
    \centering
    \includegraphics[width=0.49\linewidth]{ 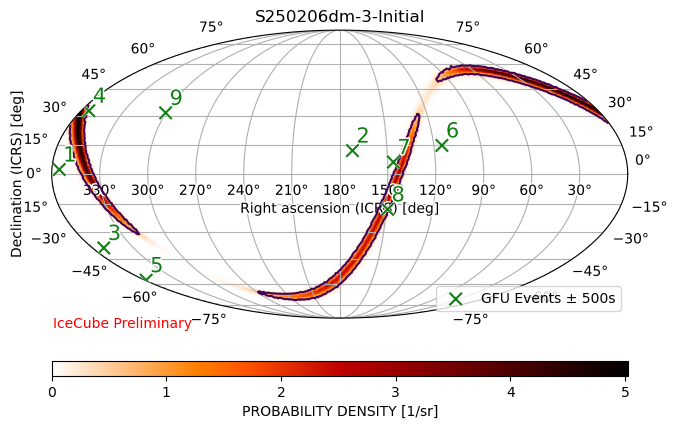}
    \includegraphics[width=0.49\linewidth]{ 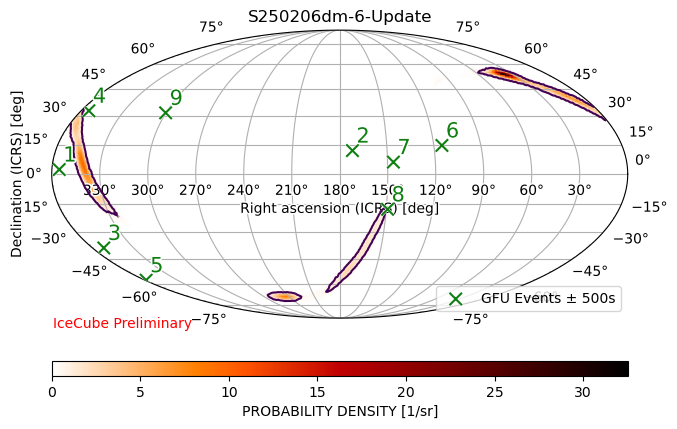}
    \includegraphics[width=0.49\linewidth]{ 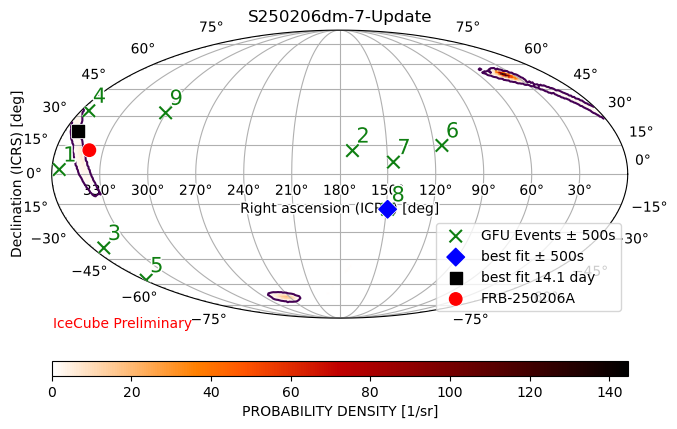}
    \caption{ Illustration of the evolution of the S250206dm (55\% NSBH/37\% BNS/8\% Terr.) skymap in real time. With the 3-Initial skymap, UML found a $p$-value of 0.008 and LLAMA 0.011. For LLAMA, neutrinos 4 and 8 reached the threshold of 10\% for release. The most significant $p$-value (0.008) from UML corresponded to neutrino 8, which prompted the team to release a GCN Circular. At the time of the GW event, Virgo was being brought online and not yet in operation mode, however approximately 1 day after the event it was determined that Virgo data could be used to refine the localization, leading to a smaller localization map (\gcn{39184}). With this updated map (7-Update), the best-fit neutrino was no longer in the 90\% contour of the GW skymap, which led to reduced significance in the analysis. The evolution of the skymap is shown for 3 different contours sent in realtime: 3-Initial, 6-Update, and 7-Update (most recent). }
    \label{fig:s250206dm}
\end{figure}

A recent event, S250206dm, from O4c received significant attention from the follow-up community as it had a 55\%  probability of being an NSBH and a 37\% probability of being a BNS merger, and if astrophysical, it has a high probability of containing a neutron star (\texttt{HasNS}$>0.99$, \gcn{39231}), making it more likely to have counterparts in other messengers. IceCube released~(\gcn{39176}) two coincident neutrinos from two different regions of the skymap~(see Figure~\ref{fig:s250206dm}). Both candidates were followed up by the DDOTI collaboration (\gcn{39198}) and Virtual Telescope Project (\gcn{39204}), both of which set upper limits on possible transient sources. Figure~\ref{fig:s250206dm} also illustrates the IceCube real-time GW follow-up program ``in action'' using three of the total seven LVK alert maps for this event. All results from the program were produced and distributed automatically via GCN Notices, promptly updating the electromagnetic follow-up community about the changes in joint significances with each skymap update. 

In addition, the S250206dm event was analyzed with the extended time window search using the UML pipeline. This search found a $p$-value of 0.85 with the most updated skymap (7-Update), consistent with no significant emission, and these results were sent as a GCN Circular (\gcn{39428}). The updated $p$-values for both pipelines in the $\pm500$ second time window were also sent in the same circular. The analysis best-fit points for the two time windows using the UML search are shown in the 7-Update skymap panel of Figure \ref{fig:s250206dm}.

\section{Comprehensive Analysis Results}

From the beginning of the O4 run until April 1 of this year, the team responded to 208 significant and 2292 low-significance gravitational wave events, and with updates released over 5325 Notices. 

\begin{figure}
    \centering
    \includegraphics[width=0.5\linewidth]{ 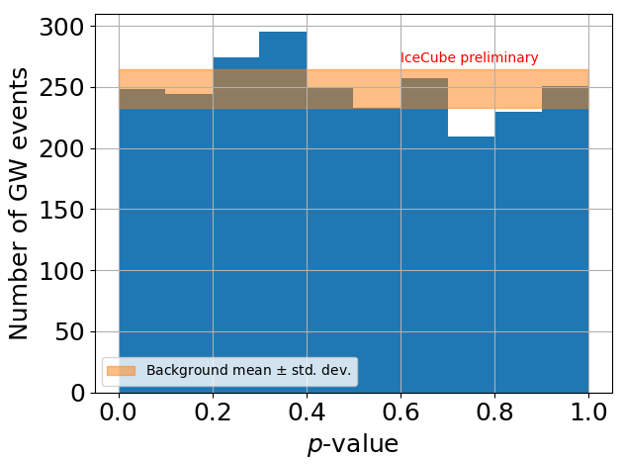}
    \caption{$p$-value distribution of the coincidences analyzed by the LLAMA pipeline in O4 until the April 2025 break. The orange band shows the expected uniform distribution and the standard deviation per bin. }
    \label{fig:llama_pvals}
\end{figure}

The LLAMA pipeline, during O4, runs in ``subthreshold mode'' and uses released information from the CBC searches in its statistical inference calculation. The $p_{\rm astro}$ attribute for each CBC event provides information on the terrestrial probability for each GW candidate, and thus, no hard cutoff is used, beyond LVK's release threshold (2/day per analysis pipeline). In Figure~\ref{fig:llama_pvals}, we show the distribution of the $p$-value results for the latest version of each event skymap from the LLAMA analysis from the first two years of the O4 run. The $p$-values in the figure were evaluated using an updated background distribution that uses sky localization information from the O4 set. The results show no significant deviation from the expected uniform distribution. A test of whether the population of GW events from the first two years of O4 emit neutrinos gives a p-value of 0.15.

\begin{figure}
    \centering
    \includegraphics[width=0.49\linewidth]{ 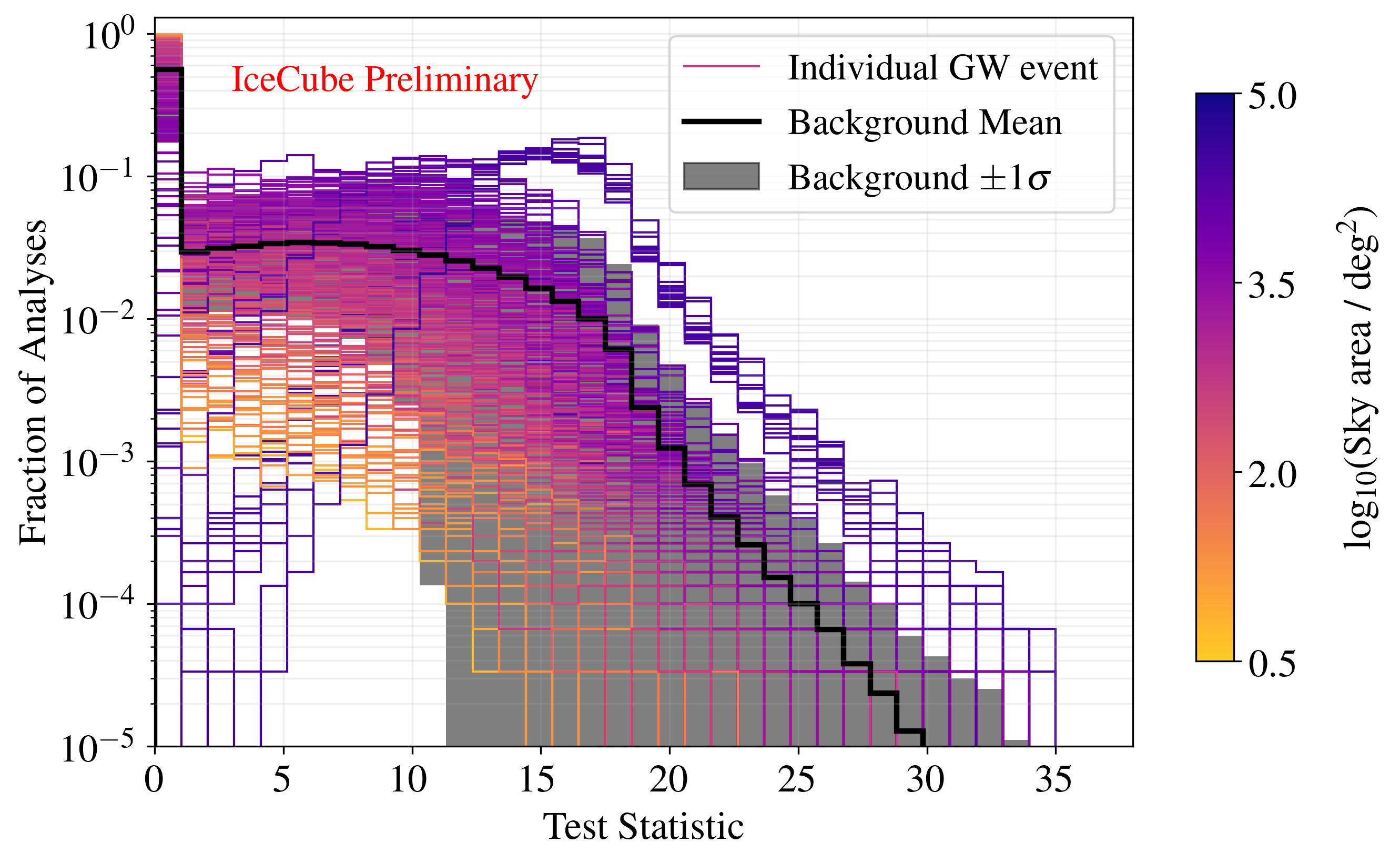}
    \includegraphics[width=0.49\linewidth]{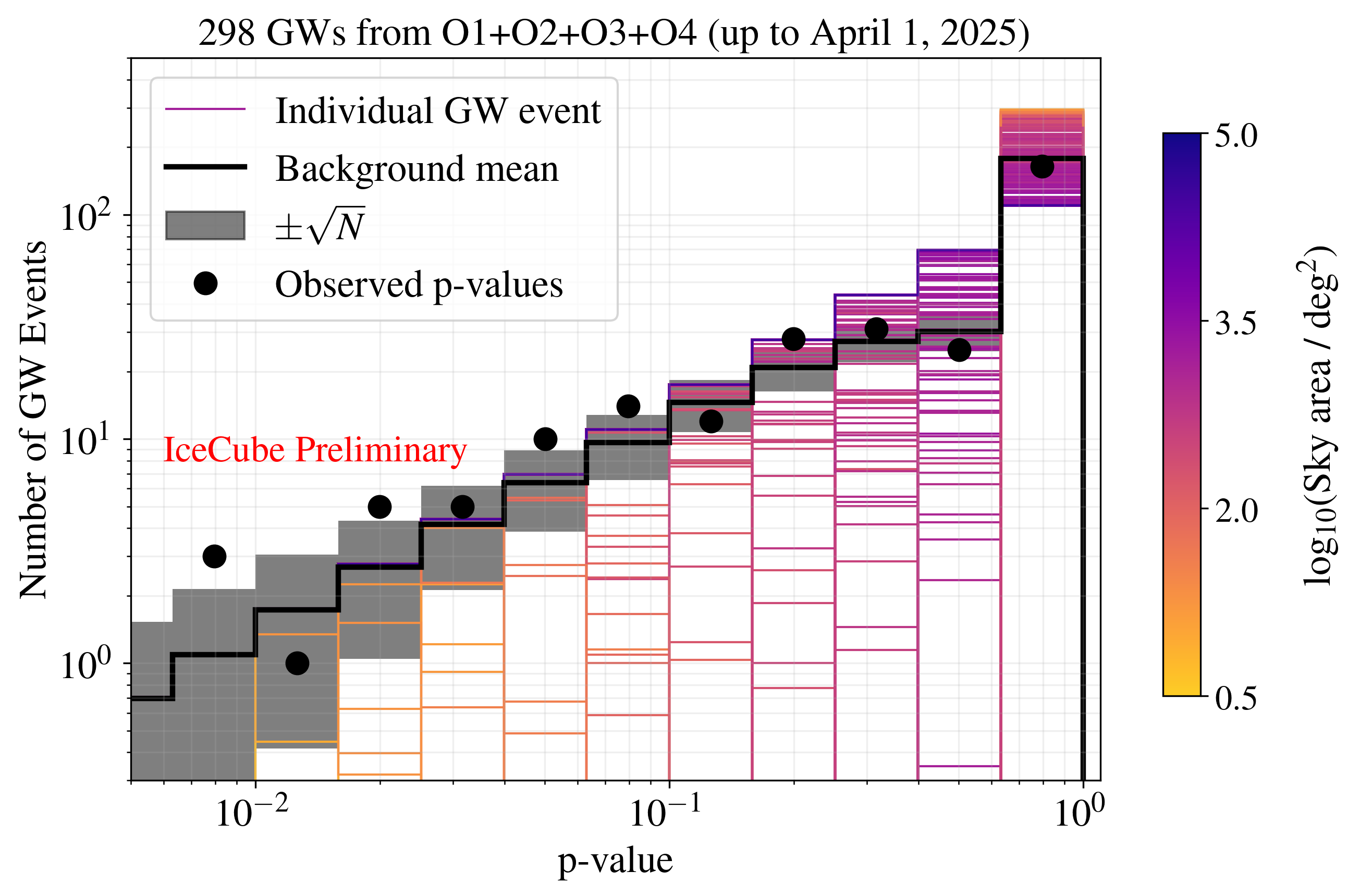}
    \caption{Background TS and p-value distributions for the unbinned maximum likelihood search for GW events in O1-O4 through April 1, 2025.
    (Left) The background-only TS distribution for each GW event (colored histograms). The area of the skymap (color bar) affects how many background-only pseudo-experiments are in the TS=0 bin. The mean (standard deviation) for each bin is shown in the black histogram (gray band).
    (Right) $p$-value histogram for the UML search. Observed p-values are shown as black points in each bin, and the background expectation (Poisson error) is shown in the black histogram (gray band). The background-only p-value expectation for each individual GW event is shown in each colored histogram.
    }
    \label{fig:uml_pvals}
\end{figure}

The $p$-value and background-only test statistic distributions for the UML search can be seen in Figure \ref{fig:uml_pvals}. The peak at $\mathrm{TS}=0 \,(p=1.0)$ in these plots corresponds to no neutrino events coincident with the GW skymap. GW skymaps with smaller skymap areas have a lower background coincidence rate from background events in the dataset, meaning more background-only pseudo-experiments have TS=0. This affects the shape of the p-value histograms shown in the right panel of Fig. \ref{fig:uml_pvals}. The background p-value expectation for the full ensemble of GW events is then the mean of the expected p-value distribution for each GW map, scaled by the number of observed GW events. This causes a non-uniform background expectation, as the skymap area for each GW event affects the shape of the per-event background p-value distribution.

\section{Conclusion and Future Prospects}

The IceCube Collaboration has been releasing coincident neutrino information in real-time since 2017 \cite{2019arXiv190105486C,stefthesis} to encourage electromagnetic follow-up of joint candidates. Since the start of the O3 run in 2019, two independent searches have been running: an analysis where a Bayesian odds ratio including astrophysical priors is used as a test statistic and a generic one using an unbinned maximum likelihood method. For the ongoing O4 observation run of LVK, both searches have been updated. A major update for LLAMA, the Bayesian method, is its new capability of performing the search and providing multimessenger statistical significance for subthreshold gravitational wave candidate events beyond the significant ones. The UML search has been automated for O4, and combined with the IceCube Fast Response Analysis, which shares the same maximum likelihood method used here \cite{IceCube_fra}.

Another significant improvement for the follow-up program is the capability to release automatic GCN Notices, which significantly decreased IceCube's real-time response time to gravitational wave alerts, and also enables prompt electromagnetic follow-up by telescopes capable of receiving automatic GCN notices. While no joint emitters of gravitational waves and high-energy neutrinos have been detected to date, the team released event information for 22 joint event candidates with $p$-values below $1\%$ to encourage electromagnetic follow-up, as demonstrated by the two optical follow-ups detailed here.
In addition, there is ongoing work to provide skymaps with joint GW and neutrino localizations for significant coincidences to the community, to increase the utility of the results produced by IceCube. 

On June 11, 2025, the LVK detectors re-started to continue their fourth observing run. While a joint detection of gravitational waves and neutrinos is not guaranteed, new data will trigger new observations and new opportunities.

\bibliographystyle{ICRC}
\bibliography{references}

\clearpage
\section*{Full Author List: IceCube Collaboration}

\scriptsize
\noindent
R. Abbasi$^{16}$,
M. Ackermann$^{63}$,
J. Adams$^{17}$,
S. K. Agarwalla$^{39,\: {\rm a}}$,
J. A. Aguilar$^{10}$,
M. Ahlers$^{21}$,
J.M. Alameddine$^{22}$,
S. Ali$^{35}$,
N. M. Amin$^{43}$,
K. Andeen$^{41}$
% ...
\\
\\
% ...
$^{10}$ Universit{\'e} Libre de Bruxelles, Science Faculty CP230, B-1050 Brussels, Belgium \\
% ...
$^{16}$ Department of Physics, Loyola University Chicago, Chicago, IL 60660, USA \\
$^{17}$ Dept. of Physics and Astronomy, University of Canterbury, Private Bag 4800, Christchurch, New Zealand \\
% ...
$^{21}$ Niels Bohr Institute, University of Copenhagen, DK-2100 Copenhagen, Denmark \\
$^{22}$ Dept. of Physics, TU Dortmund University, D-44221 Dortmund, Germany \\
% ...
$^{35}$ Dept. of Physics and Astronomy, University of Kansas, Lawrence, KS 66045, USA \\
% ...
$^{39}$ Dept. of Physics and Wisconsin IceCube Particle Astrophysics Center, University of Wisconsin{\textemdash}Madison, Madison, WI 53706, USA \\
% ...
$^{41}$ Department of Physics, Marquette University, Milwaukee, WI 53201, USA \\
% ...
$^{43}$ Bartol Research Institute and Dept. of Physics and Astronomy, University of Delaware, Newark, DE 19716, USA \\
% ...
$^{63}$ Deutsches Elektronen-Synchrotron DESY, Platanenallee 6, D-15738 Zeuthen, Germany \\
$^{\rm a}$ also at Institute of Physics, Sachivalaya Marg, Sainik School Post, Bhubaneswar 751005, India \\
% ...

\subsection*{Acknowledgments}

\noindent
The authors gratefully acknowledge the support from the following agencies and institutions:
USA {\textendash} U.S. National Science Foundation-Office of Polar Programs,
U.S. National Science Foundation-Physics Division,
U.S. National Science Foundation-EPSCoR,
U.S. National Science Foundation-Office of Advanced Cyberinfrastructure,
Wisconsin Alumni Research Foundation,
Center for High Throughput Computing (CHTC) at the University of Wisconsin{\textendash}Madison,
Open Science Grid (OSG),
Partnership to Advance Throughput Computing (PATh),
Advanced Cyberinfrastructure Coordination Ecosystem: Services {\&} Support (ACCESS),
Frontera and Ranch computing project at the Texas Advanced Computing Center,
U.S. Department of Energy-National Energy Research Scientific Computing Center,
Particle astrophysics research computing center at the University of Maryland,
Institute for Cyber-Enabled Research at Michigan State University,
Astroparticle physics computational facility at Marquette University,
NVIDIA Corporation,
and Google Cloud Platform;
Belgium {\textendash} Funds for Scientific Research (FRS-FNRS and FWO),
FWO Odysseus and Big Science programmes,
and Belgian Federal Science Policy Office (Belspo);
Germany {\textendash} Bundesministerium f{\"u}r Forschung, Technologie und Raumfahrt (BMFTR),
Deutsche Forschungsgemeinschaft (DFG),
Helmholtz Alliance for Astroparticle Physics (HAP),
Initiative and Networking Fund of the Helmholtz Association,
Deutsches Elektronen Synchrotron (DESY),
and High Performance Computing cluster of the RWTH Aachen;
Sweden {\textendash} Swedish Research Council,
Swedish Polar Research Secretariat,
Swedish National Infrastructure for Computing (SNIC),
and Knut and Alice Wallenberg Foundation;
European Union {\textendash} EGI Advanced Computing for research;
Australia {\textendash} Australian Research Council;
Canada {\textendash} Natural Sciences and Engineering Research Council of Canada,
Calcul Qu{\'e}bec, Compute Ontario, Canada Foundation for Innovation, WestGrid, and Digital Research Alliance of Canada;
Denmark {\textendash} Villum Fonden, Carlsberg Foundation, and European Commission;
New Zealand {\textendash} Marsden Fund;
Japan {\textendash} Japan Society for Promotion of Science (JSPS)
and Institute for Global Prominent Research (IGPR) of Chiba University;
Korea {\textendash} National Research Foundation of Korea (NRF);
Switzerland {\textendash} Swiss National Science Foundation (SNSF).
\end{document}